\documentclass{article}
\usepackage[utf8]{inputenc}
\usepackage[yyyymmdd]{datetime}
\usepackage{cite}
\usepackage{url}
\usepackage{tabularx}
\usepackage{multicol}
\usepackage{tabto}
\usepackage{tikz}
\usepackage{wasysym}
\usepackage{flowchart}
\usepackage{breakurl}
\usepackage{booktabs}
\usepackage{tfrupee}
\usetikzlibrary{shapes,arrows.meta,chains}

\title{Digital currency hardware wallets and the essence of money}
\author{Geoffrey Goodell\\University College London\\\texttt{g.goodell@ucl.ac.uk}}
\date{\small \textit{This Version: \today}}

\newcommand{\cz}[1]{\textit{\textbf{#1}}}

\newcolumntype{L}[1]{>{\raggedright\arraybackslash}p{#1}}
\newcolumntype{C}[1]{>{\centering\arraybackslash}p{#1}}
\newcolumntype{R}[1]{>{\raggedleft\arraybackslash}p{#1}}

\begin{document}

\maketitle

\begin{abstract}

Many proposals for the design and implementation of digital wallets assume that
the purpose of the wallet is to enable offline payments via custodial accounts,
ignoring the real problems faced by individuals and businesses that engage in
retail payments, such as the anticompetitive behaviour of payment platforms and
the decline of cash.  More importantly, the proposals ignore the \textit{raison
d'\^etre} of digital currency as a kind of digital money that can be held
independently of custodians.  Finally, the proposals demonstrate a profound
lack of imagination about the nature of digital money and the devices that
could be used to hold, manage, and exchange it.  From these presumptions flows
a set of architectural requirements that stifle the promise of digital currency
to deliver novel and efficient ways to exchange value in the digital economy.
In this article, we critically assess the essential problems that digital
currency solutions are being proposed to solve, particularly with respect to
the future of payments and the future of cash.  We evaluate the validity of
common justifications for account-based payments and certified hardware in the
context of alternative designs, limitations, and trade-offs.  We conclude,
referencing some specific designs, that the interests of consumers would be
better served by design approaches to digital currency that anticipate that
digital assets would be held outside accounts, stored offline, but transacted
online, without requiring the use of trusted hardware.

\end{abstract}

\section{Introduction}

The proliferation of cashless payments has given rise to a growing number of
merchants who refuse to accept cash, a well-documented and legal practice in
some jurisdictions around the world~\cite{dnb2023,grabar2018} that has
correspondingly given rise to government reports and
consultations~\cite{gb2025,au2024}.  Combined with the emergence of novel
private-sector payment systems, the decline of cash subverts the authority of
central banks to implement monetary policy.  This development has, in turn,
focussed the attention of central banks and other financial institutions around
the world on digital currency, particularly the idea that central banks could
issue \textit{central bank digital currency} (CBDC) for retail use.  In
principle, retail CBDC would be a direct obligation of the central bank, like
cash or central bank reserves, but existing in digital form and held by
ordinary consumers and non-financial businesses.

Some proponents of retail CBDC, including central banks, have argued that it
should have cash-like properties to deliver value to
end-users~\cite{ecb2021,boe2023}.  This argument is at least partially right.
For example, it might be expected that a cash-like system would afford users
privacy by design, resistance to discrimination and profiling, and the ability
to possess and control the monetary assets that they use to participate in the
economy and satisfy their economic needs.  These are certainly desirable
properties.  However, some authorities have interpreted the argument that a
digital payment system must be cash-like to mean that a digital payment system
must enable ``dual-offline'' transactions, namely those without connectivity of
any kind between either of the transacting parties and any third parties, in
contrast to ``single-offline'' transactions, wherein a mutually trusted third
party is reachable by one of the counterparties.  The ``dual-offline'' argument
constitutes a leap of logic that must be examined carefully, starting with its
implicit assumptions that any new payment system must be appropriate for all
scenarios, that existing solutions such as cash are insufficient to handle
payment scenarios in which neither party has access to a network connection,
and that there is a specific need for future digital currency infrastructure to
enable a kind of digital transaction that is not commonly available to users
today.

Importantly, the idea that digital currency transactions must mirror the
current system for electronic retail payments demonstrates an assumption that
all digital money must take the form of an account offered by some service
provider who acts as a custodian, and that the main promise of digital currency
concerns accessing such an account for use in dual-offline payments (an
``offline account'').  Following this logic, a digital currency hardware wallet
would be a tool by which an authorised account holder could conduct
peer-to-peer transactions between devices in the absence of network
connections.  To achieve this, the wallet itself must enforce a set of rules,
for example, the wallet must track the size of the account balance, and the
amount of money that the account holder can spend must be limited to that
value.

However, there is another, competing view that maintains that the purpose of
digital currency is to provide a digital form of cash (``e-cash''), and this
view is supported by past digital currency design proposals over the past four
decades.  Some of the most groundbreaking systems, including not only
Bitcoin~\cite{bitcoin} but also DigiCash~\cite{digicash} two decades earlier,
drew attention by introducing a way to hold money outside of custodial
accounts.  Both of these systems were sometimes described as ``cryptocurrency''
because they were engineered using cryptographic techniques.  Importantly, the
tokens that users held were explicitly designed to be cash-like, as can be
inferred from their names, and intended to be used for payments and not linked
to accounts.  In the case of Bitcoin, tokens were created and spent using an
immutable ledger (in the form of a blockchain system), and in the case of
DigiCash, tokens were created and spent as bearer instruments using a
redemption process with a central operator, which is assumed to be responsible
and accountable for its issuances, which can be held anonymously by prospective
payers.  The idea that central banks would never support such an architecture
ignores the fact that central banks have always supported the issuance and
circulation of tokens outside accounts, in the form of cash.  Within this
framework, offering cash in digital form is the next logical step.  We suggest
that the ``e-cash'' vision of digital currency is largely absent from the
discussion of the design of digital currency hardware wallets, and we argue
that this neglect is a mistake.  Contemporary examples of ``e-cash'' systems
with significant deployments include GNU Taler~\cite{taler} and (\textit{de
facto}) Privacy Pass~\cite{privacypass}, both of which feature the use of
non-custodial wallets to hold tokens for redemption.

In both the ``offline account'' and ``e-cash'' use cases, dual-offline
transactions impose additional security requirements on the transacting
parties, in addition to the security requirements applicable to digital
transactions in general.  In particular, payers must be prevented from falsely
claiming that the money they are using to pay has not previously been spent (a
``double-spending attack'').  Some prominent experts have proposed ``trusted
computing''~\cite{tcg2004,trusted-computing} as a solution to the security
problem, reasoning that if the possibility that a payer (or a payer's device)
might  misbehave, then perhaps there is benefit to be realised in making an
effort to ensure that the devices that can hold them must be able to enforce
rules directly, a ``hardware root of trust''.  After all, many users are
accustomed to carrying certified hardware, and the prevalence of certified
hardware among common user devices might be seen as a justification for their
use in the context of digital currency.  For example, smart cards commonly rely
upon tamper-resistant chips containing secrets issued by their
manufacturers~\cite{smart-cards}.  Mobile phones often contain secure enclaves
to restrict what their users can do with their devices, for example to prevent
users from copying multimedia files.

E-cash systems are designed to work without tracking the ownership of
individual tokens on a ledger.  A generic asset ``template'' containing
information such as a public key is first created by a prospective consumer,
who then uses a zero-knowledge protocol to request a privacy-preserving
certification of the digital asset, such as a blind signature, from the issuer,
which then becomes part of the asset.  The issuer can now recognise the
certified digital asset as valid without recognising its owner or the
circumstances of its creation.  The asset is stored by the end-user directly
and is not recorded on a ledger.  Double-spending can be prevented in a few
ways, one of which is to assume (a) that the issuer processes all transactions
as redemptions, (b) that the issuer maintains a ledger to keep track of the set
of redeemed assets that it receives in these transactions, and (c) that the
issuer, upon receiving an asset in a transaction, checks the ledger to verify
that the asset has not already been redeemed~\cite{chaum1982,chaum2021}.  An
alternative approach, which we call \textit{oblivious management}, does not
require the issuer to process transactions or maintain a ledger of redeemed
tokens.  In such a system, the assets carry their own provenance information,
linking transaction information they carry to rolled-up aggregation entries,
one per time period, on a ledger~\cite{goodell2022,friolo2025}; however,
neither the tokens nor the value they represent are represented on the ledger
at all.  In both cases, issuers might keep accounting records of the tokens
that they issue or redeem, but the privacy-preserving certification unlinks the
issuance of an asset, where the party requesting a new asset may be identified,
from its subsequent payment transaction and redemption, wherein only the payee
is known.

We also consider the conflation of \textit{technology}, which can be freely and
directly deployed by an individual or business to perform a function, with
\textit{services}, which are functions performed by third parties.  In a world
of smartphone applications for which most of the computation is nearly always
performed by remote service providers, it is easy to ignore this distinction.
However, since the ability to possess and control cash directly is functional
to its use, we must be careful to distinguish between technology whose
functions are governed by the end-user and technology whose functions are
governed by a third party.

Of course, the fact that certified hardware controlled by third parties is
commonly used in some contexts does not justify its use in others.  It
certainly does not justify mandating the use of such methods in the design of
future public infrastructure, especially if doing so would compel users to
relinquish some of their control to third parties.  It is all too easy for
those in the payments industry to assume that everyone has or should have a
smartphone and a bank account.  Mandating the use of certified hardware carries
many costs, and the benefits of certified hardware are not as clear as its
proponents claim~\cite{gide-starsign,thales-hsm}.

In particular, we suggest that much of the conversation in Europe about digital
currency wallets has followed from the expectation, perhaps inspired by the
digital euro regulation proposed by the European Commission~\cite{ec2023}, that
for an institutionally accepted digital currency in general, and central bank
digital currency in particular, the salient feature of cash is that it can be
transferred in a peer-to-peer way.  We disagree.  We argue that the salient
feature of cash is that it is a bearer instrument, and because peer-to-peer
transactions are not a necessary characteristic of bearer instruments, we argue
that it is worth considering protocols that do not rely upon wallets to enforce
the rules, but for which the locus of enforcement lies with service providers.
This argument has separately been articulated by the GNU Taler
project~\cite{grothoff2021}.  Even if such protocols rely upon at least one
party to be online, they can be considered social-welfare enhancing with
respect to protocols that rely upon users holding devices that they do not
control, particularly given the well-established vulnerabilities in trusted
hardware~\cite{taler2021}, and notwithstanding assertions offered by the ECB
that some hardware technology, such as embedded secure elements and SIM
devices, are more robust against attacks than trusted execution
environments~\cite{ecb2025}.

The aforementioned considerations raise the question of how end-users will hold
digital currency assets, and in particular, what the role of a \textit{wallet}
should be.  ISO 22739:2024 defines a wallet as an ``application or mechanism
used to generate, manage, store or use private keys and public keys or other
digital assets''~\cite{iso22739:100}.  The simplicity of this definition would
seem to understate the potentially complex features required to implement token
management and custody, an exploration of which is given by Barresi and
Zatti~\cite{barresi2023}.  We address some of the assumptions that Barresi and
Zatti make about the design of wallets and suggest different ways of viewing
the problem.  In particular, we critically address the arguments that
dual-offline payments constitute the most important use case for digital
currency (see Section~\ref{s:offline}), that trusted computing is an
indispensable enabler of digital currency solutions (see
Section~\ref{s:perils}), that digital currency must be held in accounts (see
Section~\ref{s:tokens}), and that a wallet is properly seen as ``not a place
where you can store cash [but] that helps you keep track of
it''~\cite{barresi2023} (see Section~\ref{s:alternative}).  Additionally, we
address the argument implicit to some e-cash solutions that issuance and
transaction processing must be performed by the same actor (see
Section~\ref{s:issuance}).  Throughout this article, we shall reference a set
of specific designs, including e-cash designs, such as GNU Taler, that are
based on bearer instruments, as well as a proposal for how to integrate bearer
instruments into a mechanism for using an integrity provider to transfer
digital assets obliviously, without maintaining records of the assets, without
immediately redeeming them for the value they represent, and without recording
any information about the assets or the counterparties involved in
transactions~\cite{goodell2022,friolo2025}.

The rest of the article is organised as follows.  In the next section, we
consider the arguments in favour of offline payments, toward an understanding
of precisely what is meant by the use of the term ``offline'', as well as the
specific kinds of offline payments for which trusted computing has been
proposed as a necessary solution and the underlying motivations for this
reasoning.  In Section~\ref{s:perils}, we critically examine the requirement
for trusted computing in the context of central bank digital currency,
identifying several key problems that undermine its perceived value.  In
Section~\ref{s:tokens}, we argue that token-based architectures are appropriate
for privacy and self-custody.  In Section~\ref{s:issuance}, we discuss the role
of the issuer as distinct from the trusted third parties that facilitate
transactions.  In Section~\ref{s:alternative}, we offer an alternative approach
that does not rely upon trusted computing and argue that this and similar
approaches are better-suited for retail digital currency infrastructure in the
general case.  In Section~\ref{s:conclusion}, we summarise our arguments.

\section{The sophistry of dual-offline payments}
\label{s:offline}

In addition to the ECB~\cite{ecb2025}, several prominent organisations,
including the International Monetary Fund~\cite{kiff2022} as well as other
central banks, such as the Bank of England~\cite{boe-tf-2022-06} and the Bank
of Canada~\cite{miedema2020}, have released reports suggesting that
dual-offline payments are desirable or even essential to the functionality of a
future retail CBDC system.  Such bold statements are striking, considering the
prevailing lack of consensus about what the requirements for digital currency
should be in general.  In this section, we shall critically explore the
argument (or implicit assumption) that the purpose of digital currency is to
facilitate access to custodial accounts in a dual-offline payment scenario.

\subsection{Kinds of payments}

First, we must ask: What exactly constitutes an ``offline'' payment?  We
consider the most common kinds of retail payment mechanisms and assess them in
terms of whether they are online or offline:

\begin{itemize}

\item \emph{E-commerce transactions.} Transactions that involve consumers using
personal computers or mobile devices to interact with websites require the
internet by definition.  They represent a significant and growing share of the
total value consumers spend in the economy, and naturally, they require
consumers to be online.  \textit{(An example is a consumer paying for goods or
services by inputting card details into a form on a web page.)}

\item \emph{Bank transfers.} Bank transfers are mechanisms for transferring
money from a sender's account with a sending bank to a recipient's account with
a receiving bank.  As account-holders, both the sender and the recipient must
be identified.  Bank transfers can be arranged online, or an account-holder can
arrange a bank transfer via a conversation with a bank teller or a computer
controlled by the bank.  The sending bank debits the sender's account, and the
receiving bank credits the recipient's account.  Settlement between the banks
is managed via an interbank clearing and settlement network.  In some cases,
interbank settlement networks cannot guarantee ``immediate'' settlement within
a few seconds, and, despite the emergence of new open banking architectures,
account-to-account ``push'' payments (in contrast to ``pull'' payments, such as
via card networks) have yet to see global adoption at the point of sale or for
transactions that require acceptance in real-time.  \textit{(An example is a
consumer instructing her bank to pay an invoice, either over the phone, in a
branch, or with a banking app or website.)}

\item \emph{Electronic card payments at the point of sale (POS).} Electronic
card payments provide a mechanism by which merchants can accept in-person
payments from consumers in real-time.  The consumer's bank (the
\textit{issuer}) and the merchant's bank (the card payment \textit{acquirer})
both participate in a \textit{payment network}.  The issuer provides a card to
the consumer, and the acquirer provides a POS device to connect the merchant to
the payment network.  The card serves as a form of identification: Its purpose
is to specify a particular bank account and to identify its user as the
beneficiary of that account.  To protect against unauthorised copying of the
identification credential, cards are usually designed with features that
provide tamper-resistance.  The client is not assumed to have a real-time
connection to an acquiring bank or service provider, and with contemporary ISO
8583 payments~\cite{iso8583}, the POS device provides a way for the card to
authenticate itself to the payment network.  If the authentication is
successful, then the payment not immediately settled, and an inter-bank credit
is implicit when the acquirer accepts the transaction.  When the transaction is
accepted, the acquirer has a credit against the issuer, which owes the
transaction amount to the acquirer.  In the case of a debit card transaction,
the issuing bank customer is debited immediately; in the case of a credit card
transaction, the issuing bank customer gets a credit in exchange for incurring
a debt against the issuing bank (credit card debt).  The payment network
informs the merchant that the transaction is approved, and the merchant can
then accept the payment in real-time.  Merchants pay interchange fees to
support this mechanism.  \textit{(An example is a consumer buying a cup of tea
by presenting a card to a merchant POS device.)}

\item \emph{Closed-loop payment systems at the point of sale.}  With
closed-loop payment networks, both consumers and merchants are primary
customers of a  \textit{platform operator}, which maintains accounts with
balances for all of its customers.  When a consumer pays a merchant in-person,
the platform operator debits the consumer's account and credits the merchant's
account.  For a transaction to succeed, either the consumer or the merchant
must have an active connection to the platform operator.  The consumer must
authenticate successfully to the platform operator, although this can be done
via the merchant's device, as it is with card payment networks.  Crucially,
banks are not generally involved in transactions; the platform operator manages
the process. \textit{(An example is a consumer using an Alipay or WeChat app to
purchase goods or services using e-money credits.)}

\item \emph{Cash.}  Cash comprises a set of physical tokens, each of which is a
direct obligation of a central bank or monetary authority.  Cash is held
directly by individuals and non-financial businesses as a store of value, and
it is also held by banks to service \textit{withdrawals}, wherein banks debit
accounts of customers in exchange for providing them with cash.  Cash is
\textit{fungible}, which is to say that individual cash tokens are mutually
substitutable and undifferentiated in practice.  Since the consummation of a
cash transaction is the exchange of a physical token, cash transactions between
parties are offline by definition.  Regulators generally do not have a direct
view into cash transactions, and assessment of regulatory and tax compliance
depends upon accounting and reporting.

\end{itemize}

E-commerce transactions are certainly online, and cash transactions are
certainly offline.  However, whether point of sale transactions and bank
transfers are online or offline is a matter of perspective.  For example:

\begin{itemize}

\item Is it enough for one party but not both to be connected to the Internet?

\item Is it necessary to be able to reach a particular service run by a
specific operator, or is it sufficient to be able to reach any one of a set of
operators?

\item Can a transaction to be provisionally allocated using the network but
consummated when both parties are disconnected from the network?

\end{itemize}

\begin{table}[ht]
\begin{center}
\sf\begin{tabular}{l>{\raggedright}p{2cm}>{\raggedright\arraybackslash}p{2cm}>{\raggedright\arraybackslash}p{2cm}>{\raggedright\arraybackslash}p{2cm}}\toprule
Payment type & payer online? & payee online? & bearer token? & dual-offline? \\\midrule
E-commerce & yes & yes & no & no \\
Bank transfer & yes & no & no & no \\
POS card payment & no & yes & no & no \\
POS closed-loop payment & maybe & yes & no & no \\
Cash & no & no & yes & yes \\\midrule
\textbf{E-token bearer payment} & maybe & maybe & yes & no \\
\bottomrule\end{tabular}\rm
\end{center}
\caption{Features of different payment types.}
\label{t:assumptions}

\end{table}

Notwithstanding some measure of understandable confusion and inconsistency in
the use of the term ``offline'' accompanying calls for considering offline
payments to be an essential feature of digital currency, it can be said that
some measure of online technology supports virtually all retail payments today,
excepting only payments in cash, which has seen a decline in use throughout the
world over the past decade, starting well before the pandemic of 2020--2022.
Table~\ref{t:assumptions} summarises the various types of payment, including
the proposed ``e-token bearer payment'', which is a digital online payment
using a bearer instrument that is currently not an option for most consumers.
It is a payment type like this that could be considered the next step in the
evolution of cash, \textit{cash for the digital economy}.

Like cash, this new bearer instrument would be possessed and controlled
directly by its users, as an assignable obligation of the issuer.  The issuer,
which could be a central bank, would be responsible for fulfilling this
obligation, but it would not necessarily hold collateral corresponding to the
total volume of its issuance.

With this in mind, it is reasonable to ask: What problem do proponents of
retail CBDC intend to solve, for which dual-offline payments constitute a
better solution than online payments involving assets that are held offline?
We acknowledge that there are some limited real-world use cases for
dual-offline payments, such as payments aboard transportation systems that lack
reliable network connectivity throughout the system~\cite{clark2018}.  However,
such examples are rare indeed, owing to the prevalence of network connectivity
in locations suitable for digital transactions.  Furthermore, we note that
payment scenarios similar to this example do not actually require dual-offline
payments; their requirements could also be satisfied by \textit{time-shifting},
in which a consumer purchases e-money tokens online and consummates
transactions with a predetermined merchant offline, by selectively sending to
the merchant tokens for which the consumer had already transferred control to
the merchant~\cite{goodell2022}.  (The consumer and the merchant can use a fair
exchange protocol to recover any unspent tokens later.) And so, it is fair to
ask the question implicit to the title of a report by the UK House of Lords,
``Central bank digital currencies: a solution in search of a
problem?''~\cite{lords2021}.

\subsection{Motivations for digital currency}
\label{s:motivations}

One might argue that conducting transactions using digital currency held by
custodians is, from the perspective of users and counterparties, not much
different than conducting transactions through ordinary bank accounts, with a
possible exception of a different regulatory environment for the digital
currency custodian.  The custodian is still a gatekeeper that could fail to
meet its obligations; the user is still bound to the custodian by
identification requirements and still does not hold the assets directly.  One
might also argue that digital currency is less useful than cash if it is
assumed that it can held only by depository institutions, or that individuals
cannot deposit and withdraw it from depository institutions, or if it cannot be
exchanged by depository institutions for central bank reserves.

We might reasonably conclude that \textit{non-custodial wallets}, which allow
users to hold digital assets directly, are a foundational benefit of digital
currency.  However, even if we accept that non-custodial wallets are essential
to digital currency, the idea digital currency held offline must be spent
offline is certainly false.  Users have an interest to possess and control
digital assets on devices of their choosing, rather than devices associated
with attributable accounts, regardless of the context in which they spend them,
for reasons of privacy and censorship resistance~\cite{kahn2005}, which we
shall address in Section~\ref{s:perils}.  Holding digital assets directly takes
custodians out of outbound transactions, allowing users to avoid the risk that
the custodians might extract fees, block certain (or all) transactions from
taking place, or build profiles of the behaviour of their customers by
monitoring the details of their transactions, such as time, location, and
counterparty.  Bearer wallets can be used in a variety of contexts, including
but not limited to in-person POS transactions.  Importantly, consumers can
realise the privacy and anti-censorship benefits of non-custodial wallets in a
fully online context such as e-commerce, for example, if they are using a
system that supports asymmetric privacy~\cite{tinn2024}.

Proponents of CBDC are right to acknowledge the secular decline in the use of
cash in retail consumer transactions as a primary motivator for digital
currency.  For example, in the UK, banks are closing ATMs and
branches~\cite{access}, making cash more difficult for consumers to withdraw
and for businesses to manage cash~\cite{tisher2020}.  Although court judgments
in some jurisdictions suggest that the legal tender status of banknotes implies
mandatory acceptance in all but exceptional circumstances~\cite{eulex2021},
jurisprudence in other jurisdictions only recognises legal tender as something
that must be accepted in the settlement of a debt~\cite{boe2025}, admitting the
possibility that vendors can freely choose what forms of payment they will
accept in the general case.

And so, when proponents of CBDC say that CBDC should allow ``cash-like''
transactions for retail users of money, they are foreshadowing the idea that
CBDC can provide individuals with a payment option that provides the key
affordances of cash, including direct control, privacy, and equal access
without discrimination.  Users of digital currency would hold direct
obligations of the central bank, as they do with banknotes issued by the
central bank, and, given the proper regulations, they might have reason to
believe that their digital currency would be accepted by retailers that prefer
electronic transactions.

CBDC can certainly address problems associated with the systematic decline in
the set of available options for consumers to access cash and the increasing
tendency for merchants to avoid payments in cash at the point of sale.  Users
can load non-custodial wallets with the help of a personal device with network
access, as well as via facilities provided by banks or ATM networks.  Consumers
can use non-custodial wallets in-person, to interface with point of sale
devices in the same manner that they use debit cards or other custodial payment
mechanisms today.  Such payments can be considered ``offline'' both in the
sense that they are in-person and in the sense that that the non-custodial
wallets themselves do not require Internet connections.  We refer to such
payment scenarios as \textit{single-offline}, because only one of the
counterparties to a transaction has access to third parties via a network or
other means of connectivity.  Single-offline payments include in-person EMV
card payments via POS systems and therefore constitute a significant share of
the payments that have replaced cash payments in the retail payments sector
over the past twenty years.

\subsection{Hold offline, transact online}

The assumption that transactions that are not between accounts must be offline
is wrong.  This assumption would imply that all payments that are not online
between payment accounts must be offline between peers that are disconnected
from real-time communication with any other party.  In particular, by stating
that ``digital euro users should have the choice to use the digital euro either
online or offline, or both'', the proposal by the European Central Bank (ECB)
for the digital euro~\cite{ec2023} implicitly suggests that the two models it
describes for payment, online via accounts and dual-offline between peers, are
comprehensively exhaustive, but this is simply false.  It simply ignores the
possibility that \textit{money can be held offline and transacted online}.
Specifically, it is possible that a consumer would hold digital currency tokens
in a non-custodial wallet and transfer them to a merchant with online
connectivity, in much the same way that a consumer might rely upon the online
connectivity of a POS device offered by the merchant.

Given the rise of e-commerce transactions, the tendency of merchants to refuse
cash at the point of sale, and the legitimate interest of consumers in privacy
and censorship avoidance, there is a clear case for using CBDC with
non-custodial wallets as a substitute for electronic card payments or network
platform transactions at the point of sale.  What remains is the question of
whether there exists a need to support payments between two parties such that
neither of which is connected to a network.  There is no evidence that
consumers and merchants who dislike cash are seeking alternative payment
technologies that do not make use of a network connection.  As a means of
facilitating a transaction between two parties in situations in which neither
party has a networked device, cash is remarkably effective.  Although it is
possible to imagine narrowly construed use cases for dual-offline payments,
such as transient network outages, those use cases do not justify system-level
requirements for a hardware root of trust on all end-user devices.

It is worth asking whether governments or central banks are harbouring an
intention to do away with cash entirely, or whether the presumption of such an
intention is shaping the view of what CBDC should be.  The argument that a
larger share of domestic transactions should be brought inside the scope of the
formal economy is certainly not new~\cite{fatf1996,camera2001}.  Recently, the
governments of certain countries, including India~\cite{pib2025} and
Indonesia~\cite{septiningrum2024}, have taken steps toward eliminating cash
entirely, an objective that is consistent with prevailing business-led
approaches to phase out cash transactions and encourage digital transactions in
general.  However, achieving that goal, whether it is legitimate or not, does
not require CBDC; some countries have already implemented regulations to
prevent the use of cash for large transactions or to control bank deposits
comprising large amounts of cash~\cite{fca2023}.  Moreover, few monetary
authorities or government regulators have called for the abolition of cash,
whereas many, including those of Switzerland~\cite{reuters2024}, the United
Kingdom~\cite{boe2024}, and the United States~\cite{fed2022}, have pledged to
preserve access to cash for the foreseeable future.  Notably, Sweden has rolled
back its drive toward creating a cashless society to support cash
infrastructure in a near-cashless economy, at great cost~\cite{loc2021}.  So
the idea that CBDC is mostly a justification for phasing out cash, as has been
speculated by some, is not corroborated by the facts, not yet at least.

\section{Certified hardware as a \textit{de facto} custodian}
\label{s:perils}

Design approaches based on certified hardware have featured prominently in the
recent debate concerning the deployment of large-scale digital currency systems
in general and systems to support CBDC in particular.  Certified hardware is a
cornerstone of the so-called ``trusted computing'' paradigm, wherein a user
carries a device designed to operate in a manner that respects the will of a
third party rather than the will of the user.  The justification for such
approaches is usually given in terms of the perceived need to facilitate
offline payments or to facilitate the recovery of lost digital assets (for
instance, in case of loss or theft of the hardware device in which those assets
were stored).  In this section, we examine the characteristics of trusted
computing within the context of its application to digital currency systems and
its potential impact on the power relationships between the users of devices,
the manufacturers of devices, and other powerful actors such as corporations
and the state.

Barresi and Zatti explicitly assume the use of trusted hardware to draw
transactions into the security envelope of a trusted
custodian~\cite{barresi2023}, implicitly drawing the hardware manufacturer into
the role of trusted third party.  In their model, the wallet would be a means
of transacting with money held in custodial accounts that would be reconciled
\textit{ex post}.

\subsection{Fair exchange}

If we define a \textit{fair exchange} as a transaction mechanism that ensures
that ``at the end of an exchange [between two players], either each player
receives the item it expects or neither player receives any additional
information about the other's item''~\cite{asokan1998}, then it is not possible
to have a fair exchange between two parties without the involvement of a third
party; this is a foundational truth~\cite{pagnia1999}.  With the exchange of
(unregistered) physical objects, possession of the object itself can be
sufficient to demonstrate the veracity of the exchange that has taken place.
But with digital objects the situation is more complicated.  It is possible to
have fair exchange without connection to the Internet, provided that a mutually
trusted third party is accessible locally, perhaps via a local network, as
might be possible with ad hoc networks.  In particular, it is not necessary for
transacting parties to have connectivity to the issuer of the assets being
transacted, since the mutually trusted third party might not be the issuer, and
it is not necessary for transacting parties to have connectivity to providers
of attributable accounts, such as banks, since the mutually trusted third party
might not be an account provider or custodian.  Transacting parties can provide
assurance, for example, in the form of cryptographic proofs or signatures, that
any applicable redemption and compliance obligations will be fulfilled.

Proponents of approaches requiring certified hardware have proposed embedding
into end-user devices a tamper-resistant ``secure element'' that can be used to
systematically restrict what the device can do.  Such restrictions can be used
to provide the immutable identity of the device to a third party, or to furnish
information contained within the device to third
parties~\cite{rfc9334,bela2024}.  The secure element would be impenetrable to
the user, and the user would not be able to modify its behaviour without
rendering the device unusable.  In effect, such an element effectively acts
as the third party by proxy, by embedding the will of the third party into the
operation of a device carried by a user.

\subsection{The perils of trusted hardware}
\label{ss:perils}

In the case of digital currency solutions that rely upon a hardware root of
trust, the secure element would presumably be used as an on-board authority
that would prevent double-spending by maintaining a record of transactions made
by the device and disallowing users from transacting tokens that have already
been signed over to others.  This design model in particular, and trusted
hardware in general, introduces a variety of problems that can be avoided if
support for dual-offline payments is not mandatory.  We consider some of those
problems here:

\begin{enumerate}

\item \emph{Trusted computing introduces significant security risks particular
to digital currency.} The risks of relying upon the security of a hardware
manufacturer are too great for a token-based CBDC context.  In particular, a
system design that relies upon trusted hardware components is fragile by
design.  If the trusted hardware fails to function as advertised or is
compromised by an attacker, then users could potentially transact the same
tokens an arbitrary number of times without bound.  Whereas with account-based
systems, such as credit and debit cards, the cost associated with the
compromise of a hardware device is limited to the size of the account with
which it is associated, for a privacy-respecting system in which money is not
linked to identity of the payer, the system-level costs of such an attack would
be huge, and the incentives would be tremendously enticing for attackers.  We
note that modern devices, such as debit cards, that rely upon tamper-proof
hardware for the exchange of value generally interact with
accounts~\cite{iso8583}.  If their security is compromised, then the value
available to attackers is limited by rules and operational procedures
implemented by custodians.  Without custodial accounts, such backup mechanisms
are not possible, and so, the consequences of a successful attack on the
hardware root of trust, such as one that impacts the cryptographic algorithm
implemented in hardware or the mechanism that protects it from tampering, might
not be addressable by a software upgrade, in turn requiring all users to
replace their devices.  Therefore, if both privacy and security are objectives
of a digital currency system, it is preferable to rely upon a protocol, rather
than the impermeability of end-user devices, to enforce the rules necessary to
achieve the required level of security.  Such a protocol can be used whenever
one of the parties has access to the network, and under such circumstances, it
would be better to allow the use of such a protocol rather than assume that
trusted hardware modules would not be breached and would guarantee that all
devices operate within the security envelope of a trusted custodian.  To assume
that all hardware wallets must take the latter approach because some
circumstances might require it is tantamount to overspecification at the
expense of users who would readily hold assets offline but transact them
online.  It is worth noting that when Chaum, Grothoff, and Moser argue that
``any economically producible device that stores tokens with monetary value in
an individual's possession, and enables offline transactions... will be the
target of successful forgery attacks...''~\cite{chaum2021}, they are referring
to wallets that perform dual-offline transactions, in contrast to wallets that
hold tokens offline and transact them online.  While overspecification of
device requirements might certainly appeal to device manufacturers seeking
compulsory mandates for their proprietary technology, it is important to
consider that protocols that do not rely upon a hardware root of trust can
function perfectly well outside the dual-offline context.  Such protocols do
not depend upon the imperviousness of end-user devices to attacks by those who
possess them, and such attacks tend to fall within the capabilities of
adversarial state actors and global crime syndicates.

\item \emph{Secure elements are unaccountable to their users.} Because trusted
hardware is impenetrable to users, users cannot know with certainty what their
devices are doing.  The inability to analyse the operations of a device
undermines the trust that users can and should place in the device.  In
essence, the device serves a second master that is not the user.  Users cannot
trust what they cannot verify, so they have reason to be suspicious of
unchecked promises on the part of hardware manufacturers or other privileged
guarantors~\cite{stallman2002}.  For example, tamper-proof components might
store and furnish information that can be used to identify the user or the
user's previous transactions.  By contrast, with \textit{free hardware
design}\footnote{``design that permits users to use the design (i.e., fabricate
hardware from it) and to copy and redistribute it, with or without
changes''~\cite{stallman2015}.}, skilled members of the community, unbound by
nondisclosure agreements, can conduct audits on the design of the hardware and
provide reports to the public.  Of course, users with the means to do so can
personally conduct the audits themselves or hire contractors to conduct the
audits for them.  However, it is not necessary for every user to actually run
an audit; evidence suggests that similar benefits are achieved so long as all
users \textit{can} perform audits if they choose to do
so~\cite{heydtbenjamin2006}.

\item \emph{Certified hardware discourages beneficial innovation.} Users who
cannot create their own devices cannot innovate.  Instituting a certification
requirement for hardware serves to ensure that only those with the time and
resources to pursue certification will be able to produce working devices.
Regulations that increase the cost of entry are known to decrease
entrepreneurship as well as innovation by existing
firms~\cite{kozeniauskas2018}.  Ordinary users and hobbyists will not be able
to develop their own devices, creating a barrier to public understanding the of
the core technology, encumbering public debate and consideration of alternative
designs~\cite{felten2002,samuelson2016}.  For this reason, hardware
certification requirements may implicitly promote security through obscurity,
contrary to accepted practices in security engineering~\cite{anderson2001}.

\item \emph{Certified hardware undermines competition by privileging its
manufacturers.} The requirement to use trusted hardware effectively locks the
hardware manufacturer in to a privileged position in the market, wherein
prospective manufacturers require certification by some authority (either
directly by the government or by a delegated authority such as a trade
organisation), and concomitant lobbying and regulation can restrict free entry,
increasing market concentration~\cite{gutierrez2019}.  In essence, enforcing
hardware certification transforms a market for consumer devices into something
more closely resembling public infrastructure procurement.  Furthermore, even
if certification were not required or if all manufacturers could have
certification at zero marginal cost, the fact that fabrication facilities carry
high fixed costs means that the market is likely to be highly concentrated,
with a few dominant manufacturers capturing the surplus~\cite{gao2017}.
Without a market to serve their needs directly, end-users will have no choice
but to accept the solution that the government chooses to trust.

\item \emph{A certified device is tantamount to a relationship with a
custodian.} Hardware manufacturers become gatekeepers at best, and globally
trusted third parties in the general case.  But unlike trusted third parties on
the network, users bind themselves to long-term relationships with personal
devices that contain trusted hardware; the relationships are not much different
between account-holders and custodians, wherein account-holders are subject to
satisfying the requirements of custodians.  Although in principle, this can be
regulated, implementation inexorably depends upon how custodians implement
regulatory compliance and is subject to the limits of the effectiveness of
regulation.  The cost of switching bank accounts is generally only a matter of
time and effort on the part of account-holders and nevertheless remains a
notoriously difficult problem for regulators, whereas the cost of switching
devices, such as smartphones, generally includes per-device manufacturing and
retailing cost plus taxes, as well as costs associated with physical
acquisition and disposal of the devices.  In addition, while it is possible to
implement both technical protocols and regulations that rely upon
locally-trusted third parties, economies of scale mean that hardware
manufacturers are likely to have global reach.  It is reasonable to conclude
that relationships between users and their devices featuring specific trusted
hardware will be sticky indeed.

\end{enumerate}

For these reasons, we are unconvinced that trusted hardware solutions are
appropriate as requirements for the design of tools that allow users to possess
and control their own digital assets.  At the same time, we would not suggest
that trusted hardware should be forbidden.  In some fraction of cases in which
physical cash is not an option and network connections are not available,
trusted hardware might be an alternative kind of third party instrument for
transacting parties to consider, although this possibility is not sufficient
justification to institute trusted hardware as a universal requirement.  In all
cases, both sender and recipient must willingly choose to trust the third party
to consummate the transaction, and given the final three points in the list
above, it is difficult to imagine that users would be able to exercise free
choice of whom to trust rather than be forced to accept a particular hardware
manufacturer.  Consider, for example, that existing users of secure hardware in
the form of electronic payment cards often have little say in which four-point
payment network their issuing banks choose, and even when they do, the set of
existing card payment networks remains well-entrenched.  Similarly, secure
hardware solutions have implications for interoperability: If the payer and
payee must use the same third-party hardware manufacturer, then network effects
might tend to entrench a single platform or small set of platforms.

\subsection{Standards and lobbying}

It is plain to see why a hardware manufacturer would lobby in favour of
regulations and standards that declare ``offline transactions'' to be an
essential requirement for CBDC.  But this argument is unsound: Cash is an
excellent offline payment method, and to our knowledge, few if any central
banks have suggested that the purpose of CBDC is to replace cash outright.  The
argument implicitly equivocates between a concept of ``offline'' that denotes a
transaction wherein a user is not required to carry a device with network
connectivity, versus a concept of ``offline'' that denotes a transaction
between two counterparties that physically meet in a location without third
parties or functioning network infrastructure of any kind nearby.  The former
scenario is the only one in which users commonly use certified hardware today,
for example, single-offline POS payments (e.g. Apple Pay) and network
applications that rely upon DRM (e.g. video streaming).  Even the use case for
the specialised Suica card, which is offered by Japan Rail for use at its
payment gates and uses embedded trusted hardware to facilitate payments when
connectivity to a central server is broken~\cite{suica}, could potentially
leverage locally reachable points of trust to aggregate transaction information
rather than assume that network connectivity is either all or nothing.
Moreover, the idea that cash could suddenly become unavailable for general use
in offline situations (the latter scenario) is an overly simplistic
justification for a system-level requirement.

We note that secure elements can be used for a variety of purposes, including
the prevention of exfiltration of private data from a device.  However,
requiring the use of secure elements for the purpose of creating attestations
to parties other than the user of the device (i.e.  ``remote attestations'')
that a device is operating within the security envelope of a trusted third
party, whether a hardware manufacturer, service provider, or authority, implies
a specific model for securing transactions that is not necessary for
single-offline payments.  For the reasons described above, we argue that
single-offline payments would be both more robust and more secure if their
locus of trust can be a network service, relying upon the various methods
network service providers can use to demonstrate the veracity of their claims,
rather than having no choice but to rely upon the integrity of trusted hardware
physically held by the other counterparty.

\begin{table}[ht]
\begin{center}
\sf\begin{tabular}{l>{\raggedright}p{1.5cm}>{\raggedright\arraybackslash}p{1.5cm}}\toprule
locus of trust & network service & certified hardware \\\midrule
supports secure single-offline payments & yes & yes \\
supports secure dual-offline payments & no & yes \\
requires wallets to be linked to attributed accounts & no & yes \\
allows anonymous payers (bearer instruments) & yes & no \\
allows unregistered or uncertified end-user devices & yes & no \\
devices may be built by user communities without approval & yes & no \\
devices may be fully auditable by end-users & yes & no \\
\bottomrule\end{tabular}\rm
\end{center}
\caption{Features of a network-based trust locus versus a hardware-based trust locus.}
\label{t:locus}

\end{table}

Table~\ref{t:locus} summarises the key differences between using a network
service as a root of trust and using certified hardware as root of trust.
Although certified hardware has the advantage of enabling dual-offline
payments, such payments are not actually common in the digital economy.  Secure
network services can provide the same degree of transaction assurance that
secure hardware devices can provide, with the only limitation being that they
must be reachable by one of the transacting parties.  At the same time, we have
argued that because it is feasible for adversarial state actors or organised
crime syndicates to compromise individual devices in their possession, it would
be too much of a risk for a system relying upon such devices to allow them to
spend money that is not linked to attributed accounts, because compromised
devices would be able to reuse anonymous money (bearer instruments) without
limit.  However, with a network service, there is no system-level risk if an
end-user device is compromised, since the locus of trust is elsewhere.  A
network service that provides payment finality\footnote{the state in which
obligations are discharged between transacting parties.} for bearer instruments
can be operated by a trusted third party, as in GNU Taler, or it can be
oblivious and based on public commitments, so that compromising it does not
enable double-spending~\cite{friolo2025}.

And so, for single-offline payments, trusted hardware adds nothing, but imposes
significant costs and limitations, and therefore it must not be considered a
system-level requirement.  Dual-offline transactions can certainly make use of
remote attestation if both parties agree, but such transactions should not be
seen as a general case for digital currency transactions, particularly given
pervasive connectivity and the security benefits of using protocol-based
security with an external root of trust rather than introducing a requirement
for counterparties to trust each other's devices.  For this reason, the
development of standards that recommend or require hardware wallets to use a
secure element for remote attestation are certainly problematic, while perfectly
functional wallet devices that provide better end-user affordances could be
designed to strictly facilitate single-offline payments instead.  Such devices
do not necessarily require network capabilities themselves, provided that their
transaction counterparties are online.

Standards that recommend hardware wallets to use a secure element to prevent
the exfiltration of private data might be appropriate in the case of
transactions involving smartphones, because of their complexity and exposure to
malware attacks.  However, it is not necessarily the case that all or most
digital wallets will be smartphones, despite being the ``benchmark in
user-accessible features'', as Barresi and Zatti have
argued~\cite{barresi2023}.  One reason is price: Although virtually everyone
must have a way to engage with the economy, a substantial share of the
population cannot afford a smartphone.  In 2024, the global median price of a
smartphone was US\$92.59, representing 11\% of global median per capita
income~\cite{gdip2024}.  Another reason is security: with ever more complex
operating systems and applications, smartphones are increasingly plagued by
security vulnerabilities~\cite{nist-mtc}, many of which specifically target
banking credentials~\cite{fadilpasic2025}.  For a sensitive function like
payments, it might be safer and more appropriate to use a dedicated
device~\cite{bizama2025}.

\section{Digital currency means tokens, not accounts}
\label{s:tokens}

Barresi and Zatti assume that wallets would represent end-user money as account
balances~\cite{barresi2023}.  However, for ordinary consumers and non-financial
businesses to hold digital currency directly, there must be some way for
end-users to hold digital currency without the requirement to have an
attributed account.  Accounts generally involve a relationship between an
\textit{account holder} and a fiduciary, typically a regulated financial
institution.  A salient property of accounts is that access to assets held
within accounts is determined by the identity of the account
holder~\cite{auer2020}, and the balance of an account represents an entitlement
that can be adjudicated, for example, in a court of law.  In contrast,
\textit{tokens} are assets that can be held outside of accounts.  Tokens can be
\textit{bearer instruments}, meaning that they can be used in a way that does
not reference the identity of their holders.  Payments can be held offline and
transacted online, so the user experience would be similar to an EMV card
payment except in which the payer presents digital assets rather than
instructions to debit an account.

Compared to assets held within accounts, assets that are held directly by their
holders have a different set of security requirements.  For example, holders of
assets held within accounts can rely upon various identification mechanisms,
such as PIN numbers and two-factor authentication, to protect their assets if
their cards or mobile devices are stolen, and asset custodians can also provide
assurances to account holders that if assets are spent against their wishes,
then the account holders can ask their asset custodians cover the value of
their losses.  In contrast, the bearer is responsible for bearer instruments.
Persons whose cash is lost or stolen generally have little recourse beyond law
enforcement mechanisms to recover their money, and such mechanisms often fail
to recover the value.  The potential for loss or theft exists not only
physically but also electronically.  Devices can malfunction, and devices can
be compromised.  The prevailing model for handling devices that no longer work
or have been compromised is to wipe their contents completely or to replace
them wholesale.  When such devices are associated with accounts, replacing
devices that have been demonstrated to be insecure might be operationally
expensive or cumbersome, but the actions of a compromised device are ultimately
traceable to the account with which it is associated, thus giving the account
holder ways to mitigate the effects.  Although the potential for loss or theft
of bearer instruments inexorably constitutes a risk that requires a different
kind of solution, approaches well-suited to digital technology, such as backup
copies and withdrawal records that can be used by victims of theft to
selectively de-anonymise their own tokens that had been stolen, can be robust
and effective in mitigating such risks.

Some central banks, including the ECB~\cite{ec2023}, the US Federal
Reserve~\cite{fed2022}, and the Bank of England~\cite{boe2023}, have presented
arguments that holding limits for wallets are necessary for financial stability
and to prevent bank runs.  However, as an alternative to the heavy-handed
approach of monitoring or restricting wallets, an approach of limiting CBDC
withdrawals from banks would be no less effective in preventing bank runs or
bank disintermediation than limiting cash withdrawals from banks, and the
approach of limiting bank withdrawals from banks has been used successfully in
the past, notably in Greece in 2015~\cite{apnews2018}.  In this sense, CBDC is
no different from cash.  To argue that requiring wallets to be managed by
parties other than their holders is necessary to address financial stability
problems resulting from reduced liquidity is to conflate the technical design
requirements for a wallet with economic justifications for eliminating the
ability for individuals to hold money outside accounts.  However, although some
banks and governments might see value in eliminating cash-like instruments and
might view CBDC as a vehicle for achieving that objective, such an argument is
not commonly articulated.

Importantly, limiting CBDC withdrawals can be done without requiring wallets to
implement account-like control mechanisms that undermine the will of their
users.  Barresi and Zatti suggest that wallets could perform ``behavioural
analytics'' on their users~\cite{barresi2023}; little could be more
disrespectful of a user's autonomy.  Similarly, the argument that digital
currency could be used as a way to implement ``programmable money'' that can
only be spent in pre-defined ways is a non-starter.  Such a mechanism would not
only undermine the singleness of money but also open a Pandora's box of ethical
concerns~\cite{li2025}, some of which have already been realised by the failed
cashless debit card scheme in Australia~\cite{greenacre2023}.  However, banks
could impose simple rules to limit the aggregate size of withdrawals that can
be made from their accounts into CBDC, both in a single transaction and over a
period of time, and such rules could be potentially combined with identity
information to limit the amount that an end-user could withdraw throughout the
system.  Such rules could also be combined with periodic token expiry, as a
mechanism to prevent hoarding~\cite{goodell2021}.  Periodic token expiry is
less intrusive than mechanisms that restrict or control wallets directly, and
it can be made transparent to users by implementing a mechanism for wallets to
automatically refresh any assets that are close to the expiration time.

Although a third party is necessary for fair exchange (and indeed trusted
hardware or closed-source software blobs can act as this third party \textit{de
facto}), Pagnia and G\"artner have demonstrated that the role of the third
party can be performed ``in zero-knowledge''.  Specifically, a third party can
facilitate final settlement of an asset transfer involving two transacting
parties, without having a custodial relationship with either party, without
knowing the identity of either party, without knowing how much is being
transferred, and even without knowing that tokens are being transferred at
all~\cite{pagnia1999}.  Importantly, with oblivious management, an issuer does
not rely upon a central ledger to track tokens individually, although the
issuer must trust that the system that collects transactions and produces
rolled-up aggregation entries, one per time period, will not equivocate.  This
trust can be enhanced by requiring that system to externalise its commitment to
a single history, for example by sending some or all of its aggregation entries
to the issuer, or by writing them to a public bulletin board or distributed
ledger.

The idea that digitalisation implies the use of accounts is not true in
general.   Digital tokens can indeed be held directly as data, with ownership
demonstrable by transferring it: using a single-use private key to transfer the
asset to another single-use private key.  A transaction is a cryptographically
signed mapping of that single-use key to a commitment to assign the asset to
another single-use key, and the ``proof of provenance'' offered by an integrity
provider demonstrates that the first single-use key has been irreversibly
mapped in this way.  Users will require a mechanism for generating keys, but
open source tools plus a good source of entropy are sufficient for this.  The
requirement of an account in the form of a ``wallet provider'' is not true for
cryptocurrencies in general, nor does a private key constitute a proof of
identity when it is created to be used exactly once.  It is true that many
Bitcoin users rely on service providers to manage their holdings.  However, to
argue from this that self-custody of tokens will never take off is pure
speculation.  Many people deliberately carry around an amount of cash that is
useful for retail purchases but small enough that they could lose it without
too much regret, despite the fact that losing it is forever.  They exercise a
free choice to determine how much cash to carry, without being forced via
technology to adhere to some limit.  More importantly, even those who might be
comfortable with a limit might be uncomfortable with the mechanism used to
impose the limit, particularly if it requires them to use a device controlled
by someone else (as described in Section~\ref{ss:perils}).

Digital assets can indeed be held by users without accounts of any kind, just
like cash and other physical assets.  An example of a digital asset that can be
held without an account is an e-cash token in Chaum's
model~\cite{chaum1982,chaum2021}, which requires every transaction to function
as a redemption, and it has also been shown that digital assets can be
transferred without redemption~\cite{goodell2022,friolo2025}.  Accounts for
recipients might be desirable, but they are not technically necessary.  As
described above, it has been proven that the prevention of double-spending does
not imply the existence of a traceable record of all the holdings and movements
through sequential time.  In Chaum's original proposal, an asset can be
transferred to a recipient without revealing information about the payer, and,
unlike oblivious management, double-spending is prevented by maintaining a
record of the spent tokens.  In the general case, and in the approach taken by
oblivious management, to prevent double-spending, it is sufficient to register
non-repudiable commitments from payers, which may take the form of a hash or
zero-knowledge proof that can be verified without knowledge of the
transaction~\cite{pagnia1999}, and proofs of these commitments can be delivered
along with the asset data rather than centrally
recorded~\cite{goodell2022,friolo2025}.  Oblivious management presupposes that
issuers can be held responsible for what they have issued and that proofs of
commitments that cryptographically demonstrate the validity and integrity of an
asset constitute sufficient evidence of an issuer's responsibility to honour a
redemption.  As discussed earlier, specific payment architectures that
implement these mechanisms have been proposed.

Tokens are also important for privacy.  Systems that make use of transactions
involving balances, even if the transactions themselves are
privacy-preserving~\cite{platypus,peredi}, intrinsically require end-users to
retain data produced during a transaction that can be used to prove that the
transaction was carried out according to the rules.  This means that
transactions are not independent of one another.  The property of
\textit{transaction independence} is important, because retaining data related
to past transactions introduces the risk that the user could be de-anonymised
via poor operational security or blackmailed to reveal information about those
transactions~\cite{friolo2025}.  Because it must be possible for balances to be
tracked through all of the movements of that balance, transactions in systems
that represent money as balances cannot achieve transaction independence.
However, systems that implement money as tokens are not subject to this
constraint.

Finally, the idea that digital currency intrinsically requires accounts is not
true; central banks can certainly issue tokens that exist outside of accounts,
just as they do with cash, and a token like this would be a liability on a
central bank's balance sheet, not unlike cash.

\section{Separating issuance from transaction processing}
\label{s:issuance}

Even if transactions are done with network access to a mutually trusted third
party, there is no reason to assume (a) that the third party must be the
issuer, or (b) that the issuer must be involved in transaction processing in
real-time.  Chaum's original proposal assumed both conditions~\cite{chaum1982},
but they are not requirements in general.  In particular, with oblivious
management, it is possible to register a (zero-knowledge) commitment to
transfer a digital asset without communicating directly with the issuer.  For a
transaction to have finality, assuming every transfer specifies the payee with
a unique single-use key, it is sufficient for both parties have confidence in a
mutually trusted third party, acting as a notary, to eventually post the
commitment (or an aggregation that comprehends the commitment) to an
append-only bulletin board that maps single-use keys to commitments and that is
trusted by the issuer~\cite{goodell2022}.  The append-only bulletin board can
be operated by the issuer directly, by a party that the issuer trusts, or as a
distributed ledger operated by a set of parties that the issuer trusts in
aggregate.  Provided that the operator of the bulletin board does not
equivocate, then since no two tokens can be transferred to the same single-use
key and have a valid proof of provenance, it is not necessary for the issuer to
maintain a ledger that records the registration status of individual
tokens~\cite{goodell2022}.

Even if we assume that all tokens are issued by a central issuer, such as a
central bank, it is certainly possible to use a decentralised system to conduct
transactions using those tokens, and wallets can be designed to respect this
possibility.  Trusting a central bank (or other issuing authority, such as one
that issues e-money) to issue currency and maintain its integrity is certainly
different from trusting it to process e-cash transactions.  In particular,
\textit{it is possible to use a decentralised system to transact centrally
issued money}.  Just as cash can be managed with a two-tiered approach, e-cash
can also be managed with a two-tiered approach.  If we assume, following Chaum,
that individual e-cash assets are generated for one-time use by consumers, then
a central bank can provide private-sector banks with ``genesis'' assets that
can be burned, \textit{pari passu}, by private-sector banks in the process of
creating single-use e-cash tokens for consumers.  Then, the single-use assets
can be recirculated within the system by private-sector banks indefinitely via
the same process before ultimately being redeemed with the central
bank~\cite{friolo2025}.  The use of a secure element to assure trust seems to
implicitly require the transacting parties to respect the authority of a
specific hardware manufacturer, and the power implicit to this authority might
be incompatible with the concept of decentralisation if all transacting parties
are forced to use secure elements produced by a limited set of manufacturers.
To require the use of a hardware root of trust with remote attestation
unnecessarily limits decentralisation.

The fact that assets held in offline non-custodial wallets can be transacted
online is sufficient to reject the requirement of a hardware root of trust for
security, because trusted hardware is unnecessary for online and single-offline
payments (see Table~\ref{t:locus}), and not every transaction must be treated
as a dual-offline transaction.  However, even if we accept that trusted
hardware is necessary for dual-offline payments, a key problem remains: Which
hardware manufacturer should transacting parties choose to trust?  The risk
implicit to requiring all users of a system to respect the same set of trusted
authorities is well-documented~\cite{stuxnet}.  The incentives for operators of
infrastructure, such as telecommunications providers operating under public
monopoly, to discourage beneficial innovation are also
well-documented~\cite{tran2019}.

We imagine that parties choosing to transact online could potentially choose
from a variety of trust providers based upon their needs and security models,
whereas forcing manufacturers of hardware wallets to embed hardware certified
by a specific authority, with the expectation that this authority is
universally and equally trusted by all users, may be too demanding.  We believe
that allowing users who need dual-offline payments to retain a choice of
mutually trusted hardware security solutions, such as removable eSIM devices,
is appropriate, although it might be impractical for users to carry around as
many hardware devices as there are commonly trusted third parties with their
various counterparties.  Interoperability does not solve this problem, since it
is the hardware itself that must be trusted.  Substituting consumer choice with
public infrastructure might seem like a tempting solution, although it entails
requiring consumers to relinquish some of their autonomy.

Finally, there is no reason for which the technical requirements for hardware
wallets used for digital currency in general should be inconsistent with the
technical requirements for hardware wallets used for other transferable digital
assets, including those that do not require a hardware root of trust because
they are not expected to support dual-offline payments.  Both kinds of wallets,
those that satisfy special requirements to support dual-offline payments and
those that do not, should be able to work within a digital currency system.

\section{Wallets as containers for bearer instruments}
\label{s:alternative}

\begin{table}[ht]
\begin{center}
\sf\begin{tabular}{l>{\raggedright\arraybackslash}p{7cm}}\toprule
possession & self-custody \textbf{versus} third party custody \\
ID requirement & yes (account) \textbf{versus} no (bearer) \\
proximity & in-person (POS) \textbf{versus} remote (e-commerce) \\
device authorisation & certified hardware \textbf{versus} public protocol \\
form & physical (cash) \textbf{versus} digital \\
issuer & central bank \textbf{versus} private sector entity \\
regulation & bank money \textbf{versus} e-money \\
\bottomrule\end{tabular}\rm
\end{center}
\caption{Some dimensions by which payment technologies can be described.}
\label{t:payments}

\end{table}

For a set of specific technical requirements for digital currency in general to
be credible, it must address the requirements of its various stakeholders.  One
of the unfortunate characteristics of the current debate is the diversity of
conflicting assumptions about the nature of payments and the relative
importance of different kinds of payment scenarios.  In reality, there are many
different dimensions that can be used to describe how people make payments
today: with and without custodians, locally and at a distance, with and without
electronic devices, with and without money issued by a central bank; see
Table~\ref{t:payments}.  Money is transferred with finality\footnote{as a way
of discharging obligations between transacting parties.} for a variety of reasons,
including to pay invoices, to purchase goods and services, to give donations,
to pay taxes, to repay loans, or to transfer funds within an organisation or
family.  These different scenarios imply a plethora of different costs and risk
profiles: Buying a cup of coffee is different from buying a house.  We should
accept that there will not be a single payment system or key mechanism to rule
them all, although it might be possible to develop a common framework that can
accommodate a variety of different payment systems.

The case for trusted hardware, as for other core mechanisms of proposed payment
systems, involves the requirement to ensure the \textit{integrity} of money
when counterparties lack access to a mutually trusted third party.  In
principle, trusted hardware can act as a trusted third party to enforce
system-level rules on user devices directly, by implementing rules that prevent
digital assets held in a wallet from being spent more than once.  However,
considering the security risks associated with compromised devices, it is
better to enforce such rules over the network instead, if it is possible to do
so.  This can be done, for example, by requiring the issuer to validate all
transactions, as proposed by Chaum and others~\cite{chaum1982,chaum2021}.
However, there is reason to believe the costs, risks, and public accountability
concerns associated with having a single central operator perform issuance,
settlement, and compliance functions might be too great for the CBDC use
case~\cite{goodell2021}.

A modern alternative to the centralised model involves using a distributed
ledger system to put the money itself on a blockchain, the way that UTXO
designs (e.g.  Bitcoin, Monero~\cite{monero}) and state-transition designs
(e.g.  Ethereum~\cite{ethereum}) do.  However, such approaches are unwieldy
because they require the ledger to track all of the tokens individually, and
transaction finality requires global consensus, introducing a system-level
economic externality with every transaction, exposing users to significant
management costs, and, in some cases, creating uncertainty about finality.  In
practice, most cryptocurrency users rely upon custodial accounts, negating a
foundational benefit of digital currency.

\begin{table}[ht]
\begin{center}
\sf\begin{tabular}{l>{\raggedright}p{2cm}>{\raggedright\arraybackslash}p{2cm}>{\raggedright\arraybackslash}p{2cm}}\toprule
feature & {UTXO\\(e.g. Bitcoin)} & Chaum (GNU Taler) & Oblivious management     \\\midrule
assets represented directly on a ledger when issued & yes   & no    & no        \\
assets represented directly on a ledger when spent  & yes   & yes   & no        \\
bundled functions shared across network peers       & yes   & no    & no        \\
finality assurance supported by network peers       & yes   & no    & optional  \\
issuance separate from settlement and compliance    & no    & no    & yes       \\
settlement separate from compliance                 & no    & no    & yes       \\
\bottomrule\end{tabular}\rm
\end{center}
\caption{Features of selected token-based payment systems.}
\label{t:systems}
\end{table}

And so, instead of aiming to share the responsibility for a bundled set of
issuance, settlement, and compliance enforcement functions across a set of
peers, it might be better to decompose the functions into distinct roles that
can be performed by different parties, enabling the costs and risks to be
allocated more efficiently.  Decentralised infrastructure can be used to
support and enable payment systems in a variety of ways, including by providing
non-repudiable evidence that a transaction had taken place, even if an
individual service provider might equivocate~\cite{goodell2022a}.
Table~\ref{t:systems} provides a comparative overview of the various
token-based payment systems that we have mentioned in terms of their key
features.

The digital currency design proposed by Goodell et al~\cite{goodell2022} and
further elaborated by Friolo et al~\cite{friolo2025} offers an alternative
approach to achieving scalability, replacing global consensus about the status
and ownership of individual assets, and the related costs of such consensus
common to modern cryptocurrency systems, with a combination of local trust and
a mechanism for keeping local trust providers honest.  By integrating a
fungible token created using Chaum's method that delivers privacy by design for
consumers into a non-fungible asset that can be transferred with strong
traceability following the initial creation, it enables the separation of the
costs and risks associated with the aforementioned roles that enable payments.
With this design, the third parties that provide transaction finality do not
manage tokens, the issuer is not involved in individual transactions, and
regulated entities, such as banks, can continue to provide compliance
enforcement (e.g. income monitoring for the purposes of taxation and sanctions
enforcement) as a condition of exchanging a spent token for a fresh token or
another form of money.

Oblivious management can work technically with a centralised notary, but in the
end, digital currency systems require common agreement about history, which is
the salient characteristic of a distributed ledger~\cite{iso22739}.  Agreement
about history does not imply shared knowledge of the holdings or transfers of
individual assets.  If assets maintain their own provenance information with
references to ledger state, then having an integrity provider record only the
zero-knowledge commitments of payers to their transactions, and having the
ledger record only the rolled-up aggregation of such commitments, is
sufficient.  These commitments can be aggregated in such a way that the ledger
grows at a fixed rate determined in advance, without individual transactions
introducing system-level externalities, and such that each transaction carries
zero marginal cost to the ledger operators.  Some established DLT systems, such
as Arbitrum, already aggregate transactions in this way~\cite{arbitrum2018}.
Proof that a transaction is part of a fixed-size aggregation, which, subject to
foundational cryptographic assumptions about the uniqueness of hashes, may be
comprehensive of an arbitrary number of transactions, is sufficient.

Trusted computing is not necessary to protect users from loss or theft, either.
To protect their assets, users may, at their option, choose devices that offer
security features such as encryption, passcodes, or local biometrics to protect
against theft, and they may make copies of their tokens to mitigate the risk of
accidental loss of their devices.  Users can retain copies of their tokens, and
they can share copies of the tokens with the police if the devices containing
them are lost, stolen, or compromised.  Users can also upload copies of their
tokens to non-custodial service providers, such as cloud storage services.  As
described in Section~\ref{s:tokens}, double-spending is prevented by the
requirement that every asset is bound to a specific integrity provider and a
specific single-use key for specifying the next transaction on the asset.

A much better way to address the limited set of cases in which ``fully
offline'' payments (that is, those where neither the payer nor the payee have
access to any suitable third parties or network infrastructure) are necessary
would be to support the use of cash itself where cash continues to work well,
such as in remote locations and environments with low connectivity, while
avoiding all of the aforementioned problems associated with trusted hardware.
To address the observed reduction of cash acceptance in urban
environments~\cite{mansuri2024,dnb2025} by providing digital currency, whilst
encouraging the provision of facilities for converting between cash and digital
currency at the boundary of ``unbuilt'' environments, would be an entirely
legitimate approach.  In particular, if one potential benefit of CBDC in an age
of waning cash infrastructure\footnote{Evidence for the decline of cash is
growing~\cite{access,tisher2020}, and it is acknowledged as an important risk by
central banks, including the ECB~\cite{ecpr2023}.} is the ability to
electronically transport money to remote population centres during
crises~\cite{goodell2020}, then there is no reason to require that central bank
money distributed in this manner must remain digital.  For example, service
providers in a particular locality can provide physical tokens in exchange for
CBDC, where the CBDC is transferred using one of the methods described earlier,
much as a casino can provide chips in exchange for cash or bank money.

None of this is to say that we should forbid limited groups of users from
trusting certified devices, if that is what they freely choose, to conduct
transactions without network connectivity to a third party.  Such an approach
is entirely compatible with an architecture that expects that the third party
would generally operate as a network service.  However, at a system level,
requiring all users to purchase and carry certified devices, even if some users
intend only to hold money offline so that they can transact it online,
introduces costs and risks that are entirely avoidable.  It is possible to
build a common framework that supports both online and single-offline
settlement involving assets held offline in free hardware devices and
dual-offline settlement involving assets held in certified hardware devices.
Oblivious management offers an opportunity to deliver on this promise by
allowing both online service providers and certified hardware devices to serve
as integrity providers for digital assets.

\section{Conclusion}
\label{s:conclusion}

In the end, we must ask ourselves: What problem are we really trying to solve
with digital currency?  Manifestly, the problem is the declining use of cash.
But this decline is really only present in environments with high network
connectivity, wherein dual-offline payments are not necessary.  For
environments with low network connectivity, the decline of cash has not
occurred, and cash continues to be the best solution for such environments.

Many central banks, including the Bank of England, have recently affirmed a
commitment to cash, while underscoring the risks associated with the decline of
cash use in favour of online payment schemes such as card payment networks.
This article has argued that the use cases that digital currency payments
should be targeting are cases for which at least one counterparty is already
online.  To argue for digital currencies with the premise that support for
dual-offline payments are its principal source of added value mainly supports
the view that such products are solutions in search of a problem, as the UK
House of Lords prominently suggested~\cite{lords2021}.

The purpose of digital wallets (and indeed, digital currency in general) is not
primarily about enabling transactions in which both parties are offline.  The
purpose of digital wallets is to allow users to hold digital monetary bearer
instruments directly, without a custodian.  Put differently, this is a way for
users to hold digital assets that they can exchange.  By facilitating such
payments, digital currency can protect the interests of consumers in all cases,
including fully networked ones.  If the counterparty has network connectivity
(for example, an electronic point of sale device), then transactions involving
digital wallets can be consummated without certified hardware.  Even if we
accept that fair exchange is impossible without the participation of a trusted
third party, and that the trusted third party can participate via the network
or via trusted hardware, it is important to underscore that a non-networked
digital wallet can conduct a fair exchange with a networked counterparty,
without the aid of special security mechanisms built into the wallet.

It is also a mistake to assume the primary motivation for CBDC is to allow a
government to phase out cash, or that custodial wallets are sufficient for all
scenarios except the narrow case in which neither the sender nor the receiver
have a communication line to a third party.  Part of the reason that cash is
successful is its role as a bearer instrument.  Credit systems actually
predated cash, and cash emerged as a partial substitute~\cite{graeber2011}.
What the world does not yet have is an institutionally supported version of
cash for the digital economy, and a digital currency aspiring to fill this role
will require a non-custodial wallet architecture.

Custodial wallets require network access as a rule, because consumers must be
able to use their devices to access their custodial accounts.  However, support
for non-custodial wallets does not imply a design choice between network access
and certified hardware in the general case.  In particular, single-offline
payments can make use of non-custodial wallets without certified hardware to
hold money offline and transact it online, leveraging the network connectivity
provided by POS devices presented to customers by merchants.  To deny the
salience of this use of wallets is to impede the development of a technology
that can protect the privacy and autonomy of users while limiting the
empowerment of big players in digital payments infrastructure.  Ultimately, the
digital wallets that support economic transactions should be seen not as access
devices or proxies offered by custodians, but as tools freely chosen, and even
built, by end-users to store and use digital money autonomously, on their own
terms.

\section*{Acknowledgements}

The author thanks Professor Tomaso Aste for his continued support for our
ongoing work on digital currencies and digital payment systems.  The author
thanks Christian Grothoff and Yuri Biondi for their generous and insightful
feedback on earlier drafts of this article.  The author also thanks his many
colleagues in the international standards development community for a series of
vigorous arguments that were instrumental to reaching the crux of this most
contentious debate.  The author also acknowledges the Future of Money
Initiative at University College London, the Systemic Risk Centre at the London
School of Economics, EPSRC, and the PETRAS Research Centre EP/S035362/1 for the
FIRE Project.

\makeatletter
\def\@biblabel#1{}
\makeatother

\end{document}